\documentclass[12pt]{iopart}
\usepackage[latin1]{inputenc}
\usepackage[T1]{fontenc}
\usepackage{ae}
\usepackage[automark]{scrpage2}
\usepackage{array}
\usepackage{amsfonts, amssymb, amsthm, amstext}
\usepackage{graphicx,subfigure}
\usepackage{hyperref}
\usepackage[usenames, dvipsnames]{color}

\newcommand{\prespectrometer}{pre-spec\-tro\-meter}

\begin{document}

\title[Model for Radon-induced background processes in electrostatic spectrometers]{Validation of a model for Radon-induced background processes in electrostatic spectrometers}

\author{N. Wandkowsky$^1$, G. Drexlin$^1$, F.M. Fr{\"a}nkle$^{1,2}$, F. Gl{\"u}ck$^{1,3}$, S. Groh$^1$ and S. Mertens$^{1,4}$}

\address{$^1$ KCETA, Karlsruhe Institute of Technology, 76131 Karlsruhe, Germany}
\address{$^2$ Department of Physics, University of North Carolina, Chapel Hill, NC, USA}
\address{$^3$ Research Institute for Nuclear and Particle Physics, Theory Dep., Budapest, Hungary}
\address{$^4$ Institute for Nuclear \& Particle Astrophysics, Lawrence Berkeley National Laboratory, CA, USA}

\ead{nancy.wandkowsky@kit.edu}

%==============================================================================
%==============================================================================

% to be published in Journal of Physics G????
\begin{abstract}

The Karlsruhe Tritium Neutrino (KATRIN) experiment investigating tritium $\beta$-decay close to the endpoint with unprecedented precision has stringent requirements on the background level of less than $10^{-2}$~counts per second.
Electron emission during the $\alpha$-decay of ${}^{219,220}$Rn atoms in the electrostatic spectrometers of KATRIN is a serious source of background exceeding this limit.
In this paper we compare extensive simulations of Rn-induced background to specific measurements with the KATRIN \prespectrometer{} to fully characterize the observed Rn-background rates and signatures and determine generic Rn emanation rates from the \prespectrometer{} bulk material and its vacuum components.
\end{abstract}

\titlepage
%==============================================================================
%==============================================================================

\section{Introduction}
\label{sec:intro}

The observation of flavor oscillations of atmospheric, solar, reactor and accelerator neutrinos has provided conclusive evidence for lepton mixing and non-zero neutrino masses~\cite{NeutrinoOscillation}. 
However, neutrino oscillation experiments only allow to assess the mass splittings of the three neutrino mass eigenstates, but yield no information on their absolute mass scale. 
The latter is of fundamental importance for both cosmology and particle physics~\cite{King}. 
In cosmology, relic neutrinos acting as hot dark matter could play a distinct role in the evolution of large-scale structures such as galaxies~\cite{m_nu_cosmo}. 
In particle physics, the determination of the neutrino mass scale would discriminate among different mass patterns, such as hierarchical or quasi-degenerate scenarios\cite{MassHierarchy}.

Various methods and techniques are employed at present to assess the absolute neutrino mass scale. 
Galaxy redshift surveys and observations of the cosmic microwave background provide information on the large-scale structure of the universe, from which upper limits on the sum of neutrino masses in the range from 200-600~meV have been derived~\cite{Cosmo}. 
In addition, experiments searching for neutrinoless double beta decay yield information on the so-called effective Majorana neutrino mass $m_{\beta\beta}$ with present sensitivities of $m_{\beta\beta}<200-400$~meV~\cite{EXO}.
In the future, these efforts are expected to reach sensitivities below 100~meV~\cite{0NBB}. 
However, one has to note that the interpretations of these observations and experiments with regard to the absolute neutrino mass scale continue to remain rather model-dependent.

On the other hand, the measurement of the electron energy spectrum close to the endpoint of nuclear $\beta$-decays such as ${}^{3}\mathrm{H}$ and ${}^{187}{\rm Re}$ or of the electron capture of ${}^{163}{\rm Ho}$~\cite{MAHO} provides the only direct and model-independent way to determine the absolute neutrino mass scale, relying only on the relativistic energy-momentum relation and energy conservation~\cite{Review}. 
The Troitsk and Mainz experiments studying the decay of (molecular) tritium $\mathrm{T}_{2}\;\rightarrow\;({}^{3}\mathrm{HeT})^{*}+e^{-}+\bar{\nu}_{e}$ with electrostatic spectrometers have yielded the most stringent experimental upper limits on the effective electron antineutrino mass $m_{\bar{\nu}_{e}}<2$~eV so far~\cite{Mainz}. 
The Karlsruhe Tritium Neutrino experiment (KATRIN) is a next generation, large-scale tritium $\beta$-decay experiment designed to determine $m_{\bar{\nu}_{e}}$ with a sensitivity of $200~\mathrm{meV}$ (90\% C.L.)~\cite{DesignReport}.
It is currently being assembled by an international collaboration at the Karlsruhe Institute of Technology (KIT) in Germany.

\begin{figure}[ht!]
 \centering
    \includegraphics[width=\textwidth]{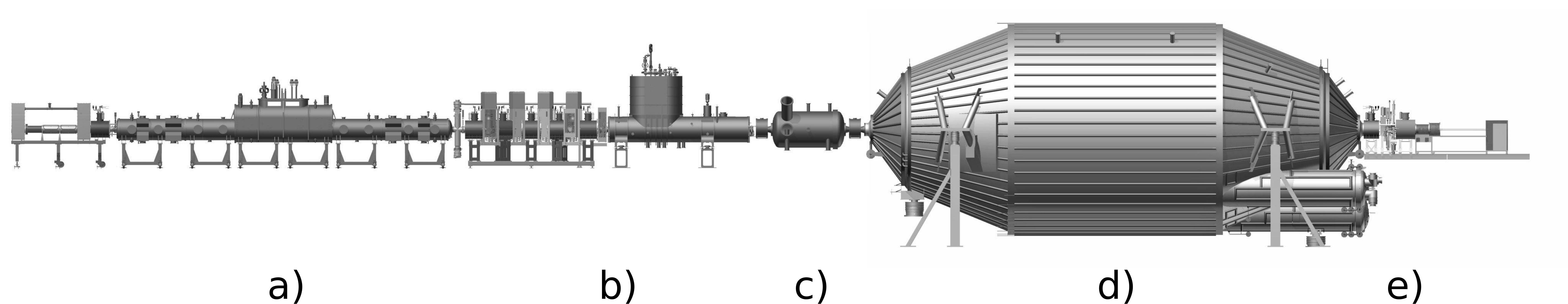}
 \caption{Overview of the KATRIN exprimental setup: a) windowless gaseous tritium source (WGTS): $\beta$-decay of molecular tritium, b) transport section: adiabatic guidance of $\beta$-electrons and removal of tritium, c) \prespectrometer{}: option of pre-filtering of $\beta$-electrons below 18~keV, d) main spectrometer: high precision $\beta$-electron energy analysis, e) detector: detection of transmitted electrons.}
 \label{fig:KATRINSetup}
\end{figure}

KATRIN will investigate the kinematics of tritium $\beta$-decay with unprecedented precision in a narrow region close to the $\beta$-decay endpoint $E_{0}\approx18.6$~keV. 
It is only in this narrow region of neutrino emission with almost vanishing neutrino momenta that one can gain access to $m_{\bar{\nu}_{e}}$. 
Figure~\ref{fig:KATRINSetup} gives an overview of the 70~m long experimental setup, which is based on a combination of an ultra-stable high luminosity tritium source~\cite{WGTS} with a spectrometer of the \nolinebreak MAC-E filter\footnote{Magnetic Adiabatic Collimation and Electrostatic filter} type~\cite{MACE,MACE1,MACE2}.
The latter is based on the magnetic adiabatic collimation of electron momenta to be analyzed by the electrostatic potential applied to the spectrometer and will be described in more detail in section~\ref{sec:MacE}. 
A segmented Si-PIN diode array allows to count the transmitted electrons as a function of the filter potential, thereby providing an integral $\beta$-decay spectrum close to $E_{0}$.
An essential pre-requisite to obtain the reference sensitivity of 200~meV is a low background level of $<10^{-2}$~counts per second (cps) in the signal region close to $E_{0}$.

In a previous publication~\cite{Fraenkle} we have reported on measurements with the KATRIN \prespectrometer{} in a test set-up configuration where $\alpha$-decays of ${}^{219,220}$Rn atoms in the volume of an electrostatic spectrometer were identified as a significant background source. 
In particular, we could demonstrate that a single radon $\alpha$-decay can produce up to several thousands of detector hits in the energy region-of-interest over an extended time period of up to several hours. 
This background results from the emission of electrons in the energy range from~eV up to several hundreds of~keV. 
The considerable range of electron energies is a consequence of the variety of processes related to the emission of the energetic $\alpha$-particle as well as the reorganization of the atomic shell.
A detailed description of these so called internal conversion, inner shell shake-off, relaxation and outer shell reorganization processes can be found in~\cite{RadonModel}.
Over almost the entire energy range, those electrons are trapped in the sensitive volume of the spectrometer due to the known magnetic bottle characteristic of a MAC-E filter~\cite{MagneticMirror,NuclearDecay}. 
Owing to the excellent ultra-high vacuum (UHV) conditions of $p<10^{-10}$~mbar~\cite{Vacuum} in the KATRIN spectrometer section, electrons remain trapped over very long periods of time, and can produce secondary electrons via ionization of residual gas molecules. 
A fraction of these secondaries can reach the detector, resulting in a background rate exceeding the KATRIN design limit of $10^{-2}$~cps.

In this paper we combine the detailed model of electron emission processes following $\alpha$-decays of the isotopes ${}^{219,220}$Rn of~\cite{RadonModel} with precise electron trajectory calculations in a MAC-E filter, which allows to describe the initial background investigations reported in~\cite{Fraenkle}, as well as the more in-depth studies performed in the course of this work and in~\cite{PhDFraenkle,PhDMertens,PhDNancy}. 
In a separate publication~\cite{NuclearDecay} we made use of the model of~\cite{RadonModel} to derive estimates of the background rates and topologies for the final KATRIN set-up, while an active background reduction technique concerning trapped electrons is described in~\cite{ECR}.

This paper is organized as follows: The field calculation and particle tracking software package \textsc{Kassiopeia}, used for our extensive Monte Carlo simulations, will be presented in section~\ref{sec:simulationtools}. Section~\ref{sec:background} then details how a single radon $\alpha$-decay can lead to a significant increase in background over a time period of up to several hours. 
In section~\ref{sec:validation}, the background model of this work will further be validated by new dedicated measurements with the KATRIN \prespectrometer{}.

%==============================================================================
%==============================================================================

\section{Simulation Tools}
\label{sec:simulationtools}

The study of event topologies of electrons from the $\alpha$-decay of ${}^{219,220}$Rn atoms, and the estimation of background rates and characteristics due to their subsequent magnetic trapping are an essential requirement in order to understand and optimize the KATRIN main spectrometer.
To meet this task, a detailed code for particle trajectory calculations in the complex electromagnetic field configuration of the KATRIN spectrometers has been developed in the frame of the code \textsc{Kassiopeia}~\cite{Kassiopeia}.
This package allows to track trapped electrons over long periods of time with machine precision. 
For the purpose of this work a Monte Carlo generator to describe electron emission following ${}^{219,220}$Rn $\alpha$-decay was developed, which is described in detail in~\cite{RadonModel}.
Therefore, section~\ref{sec:generation} will give only a short recap of the physics processes taken into account.
Section~\ref{sec:tracking} then describes the field, tracking and scattering modules of the \textsc{Kassiopeia} package, which are based on FORTRAN and C codes developed between 2000 and 2008 by one of us (F.{}~G.). The simulation software allows an extremely precise and fast computation of the relativistic motion of charged particles in electromagnetic fields. 

\subsection{Particle generation}
\label{sec:generation}

A major part of the \textsc{Kassiopeia} package is devoted to event generators for the modeling of different physical processes occurring within KATRIN. 
For the investigations of this paper, a Monte Carlo event generator was developed to describe the processes accompanying the initial radon $\alpha$-decay. 

When an $\alpha$-particle passes the fast inner atomic electrons, the direct collision process can lead to the emission of a shake-off electron~\cite{ShakeOff,KShakeOffRadon,LMShakeOffRadon}.
The resulting electron energy spectrum shows a higher-order potential dependence~\cite{Bang} because the decay energy is shared between the $\alpha$-particle and the emitted electron, which carries only a small fraction, usually of the same order of magnitude as the shell binding energy $E_{\mathrm{b}}$.

In the decay ${}^{219}\mathrm{Rn}\,\rightarrow\,{}^{215}\mathrm{Po}^{*}$, there is a probability of about 3\% for the daughter nucleus to be found in an excited state.
The inner shell electron wave function in particular can extend into the nucleus and interact with the excited state, resulting in the emission of a high-energy (up to 500~keV) internal conversion electron~\cite{ConversionDataRn219,ConversionDataRn220}.

Both processes leave vacancies in the atomic shell, which gives rise to complex relaxation cascades~\cite{Larkins,Pomplun}.
Non-radiative transitions lead to the emission of Auger or Coster-Kronig electrons.
The resulting discrete energy spectrum ranges from a few~eV up to several keV.

In the specific case of an unperturbed atomic shell or in the case of outer-shell shake-off, the shell reorganization electrons ($6p^{6}\rightarrow6p^{4}$) share an energy of 230~eV, which results in a flat energy spectrum~\cite{ShellReorganization,ShellReorg}. 
Due to their identical nuclear charge, the inner shell shake-off and shell reorganization contributions of the two polonium isotopes are assumed to be identical.

\subsection{Particle tracking}
\label{sec:tracking}

Magnetically trapped electrons from radon $\alpha$-decay with energies up to several hundred~keV have to be tracked over path lengths of several~km down to very low energies of a few~eV to fully understand their impact on background issues.
In doing so, a challenging requirement is to reach a position resolution in the $\mu$m range which corresponds to the typical cyclotron radius in a strong magnetic field region of several~T. 
In order to perform this task, the \textsc{Kassiopeia} package includes a full particle tracking module. 
The equations of motion of the electrons are being solved using Runge-Kutta methods described in~\cite{RungeKutta1,RungeKutta2,RungeKutta3}. 
Owing to the complexity of the inner electrode~\cite{Electrode} and magnet system~\cite{Aircoil} realized in the KATRIN experimental setup, the calculation of electric and magnetic fields is most challenging. 
To do so we make use of the zonal harmonic expansion~\cite{FerencEl,FerencMag}, and, in the case of electric field computations, the boundary element method~\cite{BEM}.

For the investigations of this paper, all processes resulting in an energy loss of stored particles play a major role. 
The corresponding \textsc{Kassiopeia} modules to describe electron cooling include the processes of elastic scattering, excitation and ionization of $H_{2}$ molecules (dominant residual gas species within the KATRIN main spectrometer).
The corresponding cross sections~\cite{ElasticHydrogen1,ElasticHydrogen2,BEB,BEB2}, energy loss values~\cite{ExcitationHydrogen,ExcitationHydrogen2} and scattering angles have been implemented in the scattering routine of \textsc{Kassiopeia}.

The scattering cross sections and energy losses vary significantly for different gas species. 
Correspondingly, primary electrons with identical start parameters experience different storage times and generate different numbers of secondary electrons.
Initial mass spectrometry measurements~\cite{PhDFraenkle} showed that the residual gas inside the \prespectrometer{} mainly consists of hydrogen, water and nitrogen, while argon was used within specific test measurements to increase the pressure to a desired value.
When studying electron cooling by scattering off residual gas, the ionization process is the dominant energy loss mechanism, contributing to >80\% of the total energy loss for electrons above 1~keV when scattering off hydrogen takes place. 
Hence, molecule-specific ionization cross sections and energy losses are used within the simulation (water, nitrogen~\cite{Molecules}, argon~\cite{Argon}).
In the case of elastic or excitation processes, energy losses are computed using molecular hydrogen input data, which is a sufficient approximation.
Arbitrary residual gas compositions consisting of hydrogen, water, nitrogen or argon can be defined via specific configuration files. 
The fact that electron cooling strongly depends on the residual gas pressure and composition has been used to gain insight into background processes by comparing measurements and simulations at different pressures, which will be shown in section~\ref{sec:validation}.

Another source of energy loss of magnetically trapped electrons is synchrotron radiation. 
Due to their cyclotron motion, electrons continuously emit synchrotron radiation. 
In the non-relativistic limit, the energy loss per unit time interval $\Delta t$ (in SI units) by this radiative process is given by
\begin{equation}
 \frac{\Delta E_{\perp}}{\Delta t}=\frac{4}{3}\frac{e^{4}}{m_{e}^{3}c}\cdot B^{2}\cdot E_{\perp}\approx 0.4\cdot B^{2}\cdot E_{\perp},
 \label{equ:synch}
\end{equation}
where $B$ denotes the magnetic field, $c$ the velocity of light, and $e$ and $m_{e}$ the electron charge and mass. 
To good approximation only the transversal kinetic energy component $E_{\perp}$ is reduced by this process. 
In our simulation, synchrotron energy losses within a Runge-Kutta step are determined by using the average magnetic field during the step. 
From eq.~(\ref{equ:synch}) follows that the cooling effect due to synchrotron radiation is most efficient for large transversal kinetic energies and large magnetic fields. 
At the same time, the scattering cross section decreases steeply for increasing electron kinetic energies, so that synchrotron losses dominate at higher energies. 
As an example for the standard operation mode of the KATRIN \prespectrometer{} ($p~=~10^{-10}~\mathrm{mbar}$, $B_{\mathrm{max}}~=~4.5~\mathrm{T}$, $B_{\mathrm{min}}~=~15.6~\mathrm{mT}$), we note that a stored electron with a kinetic energy of $300~\mathrm{keV}$ (with 150~keV in the transversal component) typically loses about half of its energy due to synchrotron radiation. 
This value can be decreased by increasing the residual pressure via the injection of argon, thereby increasing scattering losses.

The implementation of these processes into the \textsc{Kassiopeia} package thus allows to study background generating processes of trapped electrons from ${}^{219,220}$Rn $\alpha$-decays in great detail.

%==============================================================================
%==============================================================================

\section{Radon-induced Background within KATRIN}
\label{sec:background}

When discussing the background processes from electrons following radon $\alpha$-decays, we first briefly outline the working principle of the KATRIN spectrometers, which is based on the MAC-E filter principle (section~\ref{sec:MacE}). 
Section~\ref{sec:backgroundmechanism} then describes the trapping mechanism of electrons in a MAC-E filter and the resulting significant increase in background. 
The final section~\ref{sec:radonbackground} is devoted to the specific case of radon-induced background.

\subsection{The MAC-E filter principle}
\label{sec:MacE}

\begin{figure}[ht!]
 \centering
 \includegraphics[scale=0.09]{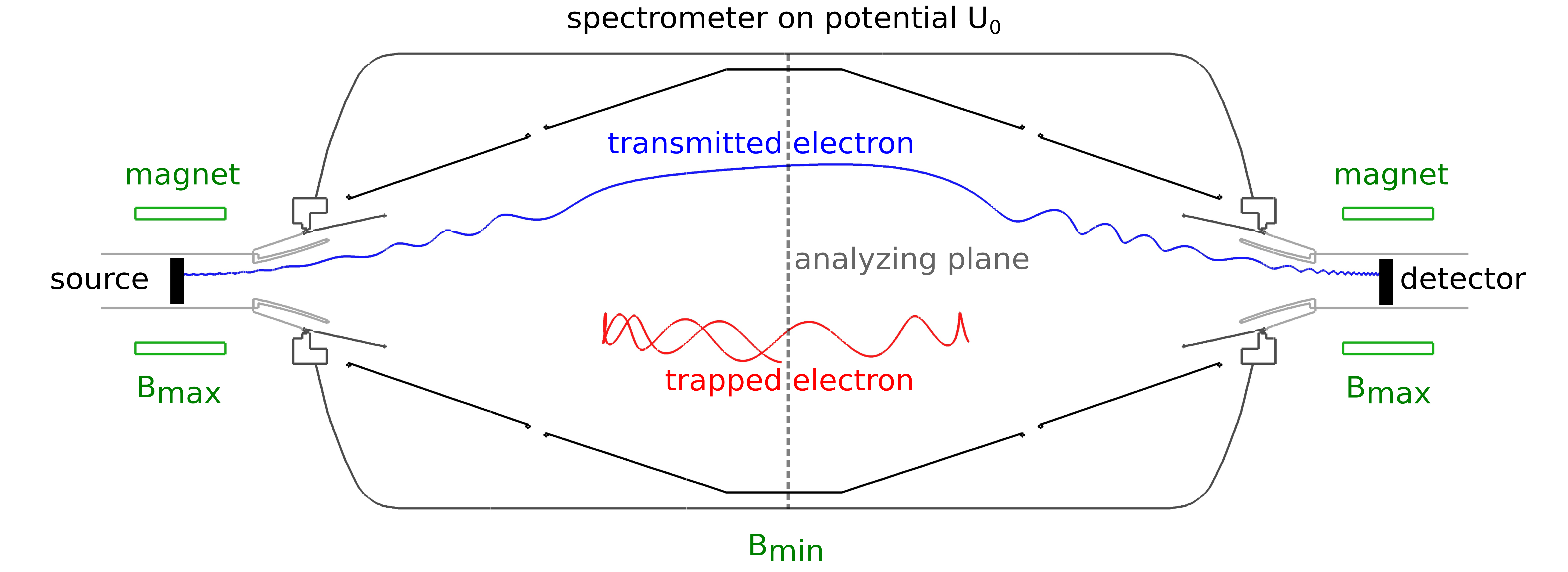}
 \caption{MAC-E filter principle. Superconducting magnets produce a magnetic guiding field. On the one hand, signal electrons, created in the source with sufficient kinetic energy, can pass the potential barrier at the analyzing plane and are counted at the detector. On the other hand, electrons generated inside the volume of the spectrometer can be trapped due to the magnetic mirror effect.}
 \label{fig:MAC-E}
\end{figure}

Electrons from tritium $\beta$-decay are emitted isotropically in the source (WGTS) and have to be guided adiabatically to the spectrometer by a magnetic guiding system which consists of a series of superconducting magnets (see fig.~\ref{fig:KATRINSetup}). 
In the spectrometer, the energy analysis takes place via the MAC-E filter technique, which is illustrated in figure~\ref{fig:MAC-E}. 
Two superconducting magnets provide a magnetic guiding field for $\beta$-electrons, while an electrostatic retarding potential $U_{0}$, applied to the spectrometer and its inner electrode system, allows to filter the signal $\beta$-electrons. 
The area where the absolute value of $U_{0}$ reaches its maximum is defined as the so-called analyzing plane. 
The kinetic energy of incoming electrons is composed of a longitudinal component $E_{\parallel}$ parallel to the magnetic field lines and a transversal component $E_{\perp}$ corresponding to the electron's cyclotron energy. 
The electrostatic potential, however, only affects and filters $E_{\parallel}$. 
Therefore, $E_{\perp}$ has to be transformed into $E_{\parallel}$ by the magnetic gradient force. 
In order to achieve this, the magnetic field strength has to decrease from its maximum value $B_{\mathrm{max}}$ at the center of the superconducting magnets to its minimum value $B_{\mathrm{min}}$ at the analyzing plane. 
The field gradient $\nabla B$, however, should not be too steep to guarantee a fully adiabatic electron motion, thereby conserving the magnetic moment $\mu~=~E_{\perp}/B$. 
This principle allows for a high-resolution energy analysis by the retarding potential. 
Electrons with sufficient kinetic energy to pass the electrostatic barrier are re-accelerated and counted at the detector, thus yielding an integral energy spectrum.

\subsection{Background production within a MAC-E filter}
\label{sec:backgroundmechanism}

While the magnetic field setup of a MAC-E filter allows for unsurpassed precision in the scanning of the tritium $\beta$-decay spectrum close to $E_ {0}$, it also acts inherently as a magnetic bottle for electrons created in the flux tube of the spectrometer (see fig.~\ref{fig:MAC-E}). 
The longitudinal energy $E_{||}$ of such an electron is transformed into transversal energy $E_{\perp}$ when propagating towards the increasing magnetic field strength at the entrance and exit region of the spectrometer. 
At the same time, the electron concurrently gains longitudinal energy by the accelerating electric potential. 
If the transversal energy of the electron is above a certain threshold, the magnetic transformation is dominant and $E_{||}$ will be converted completely into $E_{\perp}$. 
Consequently, this electron is reflected by the \textit{magnetic mirror effect}~\cite{MagneticMirror}, which results in a stable storage condition within the spectrometer volume.

When trapped, electrons scatter off residual gas species, thereby slowly cooling down until their transversal energy drops below the storage threshold and they can escape the trap. 
In the course of this process, which can take up to several hours, secondary low-energy electrons ($<100$~eV) are produced via ionizing collisions. 
These secondaries are accelerated by the retarding potential and, when escaping the magnetic bottle, will hit the detector within the energy region-of-interest, thus producing an irreducible background class.

Depending on its initial kinetic energy, a single stored primary electron can produce up to several thousands of secondary electrons which contribute to the background.
Due to its non-Poissonian nature, this background source can significantly constrain the neutrino mass sensitivity of KATRIN~\cite{NuclearDecay}, if no countermeasures~\cite{ECR} are taken. 

\subsection{Radon-induced background in the \prespectrometer{}}
\label{sec:radonbackground}

In the framework of the \prespectrometer{} electromagnetic test measurements~\cite{PhDFraenkle,Fraenkle}, a background source with the characteristics described above was identified to stem from electrons emitted during the $\alpha$-decay of single ${}^{219,220}$Rn atoms. 
While background rates close to the intrinsic detector background of $(6.3\pm0.2)\cdot 10^{-3}~\mathrm{cps}$ were observed most of the time, specific time intervals of up to two hours duration showed enhanced background rates of up to $250\cdot 10^{-3}~\mathrm{cps}$. 
These distinct intervals occurred about 7 times per day, each caused by a single nuclear decay of a specific radon isotope.

There are various potential sources of radon emanation to take into account.
First, the vessel of the \prespectrometer{} with its diameter of 1.7~m and length of 3.3~m (see fig.~\ref{fig:MAC-E}) features a large inner surface of $25~\text{m}^{2}$ including weld seams which are a potential source of radon emanation~\cite{Gerda}. 
In addition, several auxiliary devices (vacuum gauges, glass windows etc.) are attached to the vessel, which can also emanate radon atoms as a result of their primordial abundance of $^{232}\mathrm{Th}$, $^{235}\mathrm{U}$ and $^{238}\mathrm{U}$.
A major source of ${}^{219}$Rn emanation was identified to be the non-evaporable getter (NEG) material~\cite{NEG}, used as an efficient pump for hydrogen. 
Details on different sources of radon emanation can be found in~\cite{Fraenkle}.

While the $\alpha$-particle itself and the fluorescence X-rays of atomic relaxation processes do not contribute to the background, energetic electrons, which are emitted during the nuclear $\alpha$-decay (see section~\ref{sec:generation} and~\cite{RadonModel}), have a large probability to be stored inside the \prespectrometer{}.
Thus, they will lose their kinetic energy via secondary processes such as scattering or synchrotron radiation. 
If we assume that electrons cool down exclusively via scattering off molecular hydrogen, the average energy lost per produced secondary electron is $\omega\approx33$~eV. 
This value was determined with the scattering routines implemented in \textsc{Kassiopeia} and is in good agreement with the calculated value of 37~eV in~\cite{Eloss}. 

For primary electrons trapped within the \prespectrometer{} flux tube, the number of secondary electrons is influenced by several effects:
\begin{itemize}
 \item For the \prespectrometer{} field configuration detailed in fig.~\ref{fig:MAC-E}, only electrons with a kinetic energy above about $E^{\text{min}}_{\perp}~=60$~eV are stored magnetically. Hence, a high-energy (keV) electron will not transform its entire kinetic energy into secondary electrons.
 \item Electrons experience non-negligible energy losses due to synchrotron radiation. According to eq.~(\ref{equ:synch}), the synchrotron losses increase for larger electron transversal energies. 
 \item Above 100~keV starting energy, electron trapping is affected by non-adiabatic effects, resulting from the specific electromagnetic field configuration of the \prespectrometer{}. Non-adiabaticity is induced if the magnetic field changes significantly within one gyration, so that the transformation of $E_{\perp}$ into $E_{\parallel}$ and vice versa is no longer proportional to the change of the magnetic field. Consequently, the polar angle of the electron will change randomly, eventually hitting a value below the trapping threshold.
 \item Additionally, electrons with very high energies have large cyclotron radii which can lead to electrons hitting the spectrometer vessel, thus prematurely terminating the background-generating process.
\end{itemize}

\begin{figure}[ht!]
 \centering
 \includegraphics[width=0.8\textwidth]{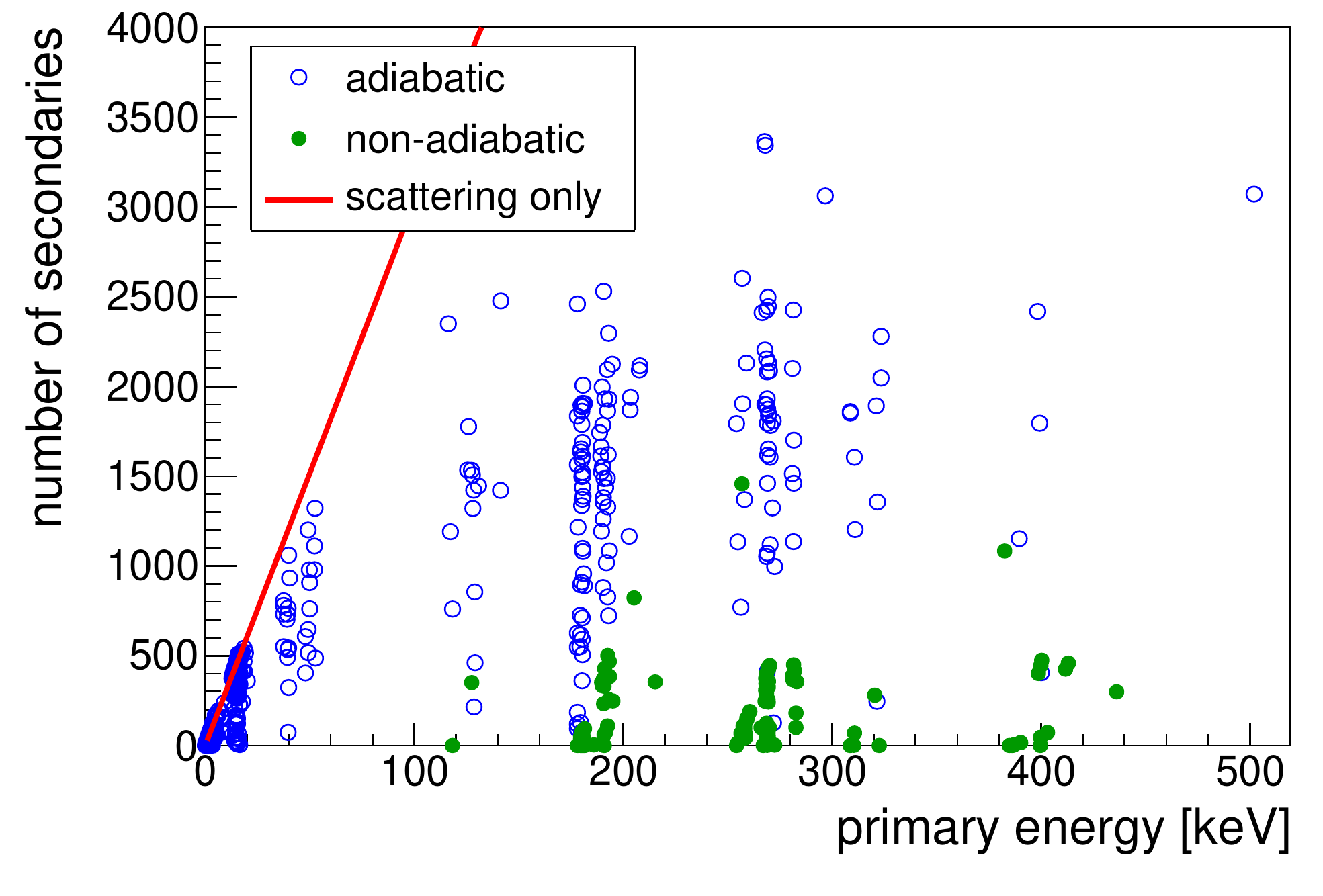}
 \caption{Number of produced secondary electrons as a function of the starting kinetic energy of the primary electron. In this case the ${}^{219}$Rn event generator was used and electrons were tracked in the \prespectrometer{}. The red line corresponds to an average energy loss of $\omega=33~\mathrm{eV}$ per ionization. The actual number of secondary electrons remains below this limit because the primary electrons also lose energy due to synchrotron radiation. The non-adiabatic events are marked as full circles.
}
 \label{fig:SecondaryRate}
\end{figure}

To study background-generating processes and non-adiabatic effects, \textsc{Kassiopeia} simulations with the ${}^{219}$Rn event generator described in~\cite{RadonModel} were performed.
A total of 10000 electrons was tracked in the \prespectrometer{} under standard operating conditions ($p~=~10^{-10}~\mathrm{mbar}$, $B_{\mathrm{max}}~=~4.5~\mathrm{T}$, $B_{\mathrm{min}}~=~15.6~\mathrm{mT}, U_{0}~\approx~-18400~\mathrm{V}$). 
Figure~\ref{fig:SecondaryRate} shows the resulting number of electrons produced via ionization processes as a function of the primary electron energy. 
From the figure it is evident that primary electrons of several hundred~keV can produce up to 3500 secondary electrons. 
Furthermore, the number of secondaries being produced by high-energy primaries is smaller than what is expected from an average energy loss of $\omega=33$~eV. 
This reduction is mainly due to the rather large synchrotron energy losses in the \prespectrometer{}. 
In addition, electrons released from the spectrometer magnetic bottle due to non-adiabatic effects contribute to this behavior.

When transferring these numbers into background rates it is important to note that only a fraction of the produced secondary electrons will actually reach the detector. 
First, only half of the electrons will escape towards the detector side of the spectrometer. 
Second, the electrostatic field configuration of this setup features a small Penning trap in the center of the \prespectrometer{} with a depth of up to 12~V in the sensitive volume. 
As a result, low-energy secondary electrons are stored within this trap with a probability of about 60~\%. 
Despite these background-reducing factors, radon-induced events can still induce enhanced background rates where up to 2000 detector hits were observed over up to 2 hours ($N_{\text{bg}}\approx280\cdot10^{-3}$~cps).

%==============================================================================
%==============================================================================

\section{Validation of Background Model}
\label{sec:validation}

In the following we report on a detailed experimental validation of our radon event generator~\cite{RadonModel} and the corresponding Monte Carlo simulations described above. 
The experimental information is based on specific measurements with the \prespectrometer{}, described in~\cite{PhDFraenkle}, motivated to give complementary high-precision information on background characteristics and mechanisms. 
In the first section~\ref{sec:measurements} an overview of the different measurements will be given. 
Section~\ref{sec:spatialdist} discusses the specific event topology of trapped electrons in the form of ring structures at the detector, which gives access to the spatial distribution of radon decays inside the spectrometer. 
A further important background characteristic is the rate of single events where Monte Carlo simulations are compared to measurements within section~\ref{sec:ratesingle}. 
In the final section~\ref{sec:activities} the Monte Carlo results are used to determine the radon activities in the \prespectrometer{} setup, which then are compared to the independent values derived in~\cite{Fraenkle}.

\subsection{Overview of \prespectrometer{} radon measurements}
\label{sec:measurements}

In order to validate our model of radon-induced background more reliably, three different \prespectrometer{} background measurements have been investigated in detail. 
A full Monte Carlo simulation of each measurement configuration has yielded consistent results, which will be presented in sections~\ref{sec:ratesingle} and~\ref{sec:activities}. 
This section first gives an overview of the measurement strategy and experimental results.

As outlined above, the storage time of an electron strongly depends on the residual gas pressure inside the spectrometer. 
Therefore, two measurements at different pressures were performed, first a measurement at the standard \prespectrometer{} operating pressure of $p_{\text{LPG}}~=~10^{-10}$~mbar with the residual gas composed mainly of hydrogen, water and nitrogen. 
Secondly, a measurement was performed at a higher pressure of $p_{\text{HPG}}~=~2\cdot10^{-9}$~mbar while injecting argon gas into the spectrometer. 
In both measurements, the vacuum system consisted of the NEG pump (emanating $^{219}$Rn) and one turbo molecular pump (TMP) for non-getterable species. 
In the following, these configurations are labeled LPG (low pressure with getter) and HPG (high pressure with getter), respectively. 
In order to definitely confirm the getter material as a major source for ${}^{219}$Rn, the NEG pump was removed for a third measurement. 
A second TMP then had to be activated to compensate for the loss in pumping power. 
As these modifications resulted in a relatively high pressure value of $p_{\text{HP}}~=~10^{-9}$~mbar, this measurement is labeled HP (high pressure - without getter). 
The number of active TMPs influences the pump-out time for gases, and thus the decay probability of the different radon isotopes, as shown in table~\ref{tab:decayProb}. 
In all cases, the decay of $^{222}$Rn can be neglected.

\begin{table}[!ht]
\caption{Pumping speeds and decay probabilities inside the \prespectrometer{} for $^{219}$Rn, $^{220}$Rn and $^{222}$Rn, depending on the number of active TMPs~\cite{Fraenkle}. The total \prespectrometer{} volume amounts to $V=8.5~\mathrm{m}^3$. }
\begin{center}
  \begin{tabular}{ccccc}
  \hline
  \hline
  \# TMPs & speed [l/s] & $^{219}$Rn & $^{220}$Rn & $^{222}$Rn \\
  \hline
  1 & 194 & 0.885 & 0.353 & $9.2\cdot10^{-5}$\\
  2 & 402 & 0.787 & 0.208 & $4.4\cdot10^{-5}$\\
  \hline
  \hline
  \end{tabular} 
\end{center}
\label{tab:decayProb}
\end{table}

The intervals with elevated background rate caused by a single radon decay were (in close analogy to~\cite{Fraenkle}) categorized into three different event classes, depending on the number of counts (cts) at the detector: CI (10-50 cts), CII (51-500 cts) and CIII (>500 cts). 
According to our simulations, CI and CII events originate from radon decays which produce conversion or shake-off electrons. 
However, only the high-energy electrons from the internal conversion process can produce CIII events. 
These conversion electrons originate practically only from the decay of ${}^{219}$Rn, with the NEG pump identified as a major source of this isotope. 
However, even after the complete de-installation of the NEG pump (measurement HP), the distinct CIII signature of ${}^{219}$Rn events was still observed, though at a greatly reduced rate. 
Consequently, we have implemented a background model where three different sources contribute to the total radon activity inside the spectrometer: ${}^{219}$Rn from the getter ($^{219}\mathrm{Rn}_{\text{G}}$), and $^{219}$Rn as well as $^{220}$Rn from the spectrometer and auxiliary equipment attached to it (${}^{219}\mathrm{Rn}_{\text{B}},{}^{220}\mathrm{Rn}_{\text{B}}$).

Table~\ref{tab:measurement} gives a summary of the measurement conditions in the three configurations, which have been used as input for our Monte Carlo simulations. 
Furthermore, the observed occurrence of CI-III events and their contributions to the total spectrometer background are given. 
These values will be compared to those derived via simulations (section~\ref{sec:activities}).

\begin{table}[!ht]
\caption{Overview of UHV measurement conditions and resulting radon-induced background rates. For the three different measurement conditions (low-pressure (LPG) and high-pressure (HPG) with getter installed, and high-pressure without getter (HP)) the events were categorized into three different classes according to the number of radon-induced counts (cts) at the detector: CI (10-50~cts), CII (51-500~cts), CIII (>500~cts). Event rates and contributions to the total spectrometer background $r_{\text{bg}}$ are shown for the individual classes.}
\begin{center}
  \begin{tabular}{lccc}
  \hline
  \hline
  measurement & LPG & HPG & HP\\
  \hline
  getter & \checkmark & \checkmark & $\times$\\
  \# TMPs & 1 & 1 & 2\\
  pressure [mbar]& $1\cdot10^{-10}$& $2\cdot10^{-9}$ & $1\cdot10^{-9}$\\
  gas composition & $H_{2},H_{2}O,N_{2}$ & $Ar$ &$H_{2},H_{2}O,N_{2}$\\
  \hline
  events/day $(CI)$ & $4.2\pm1.4$ & $5.5\pm1.2$ & $2.0\pm0.6$\\
  $r_{\text{bg}}(CI)$ [$10^{-3}$ cps] & $0.85\pm0.1$ & $1.1\pm0.1$  & $0.49\pm0.03$\\
  \hline
  events/day $(CII)$ & $1.7\pm0.9$ & $0.8\pm0.5$ & $0.24\pm0.24$\\
  $r_{\text{bg}}(CII)$ [$10^{-3}$ cps] & $3.6\pm0.15$ & $1.6\pm0.1$ & $0.17\pm0.02$ \\
  \hline
  events/day $(CIII)$ & $1.0\pm0.7$ & $0.8\pm0.5$ & $0.31\pm0.28$\\
  $r_{\text{bg}}(CIII)$ [$10^{-3}$ cps] & $8.1\pm0.2$  & $10.1\pm0.2$ & $1.38\pm0.05$\\
  \hline
  \hline
  \end{tabular} 
\end{center}
\label{tab:measurement}
\end{table}

\subsection{Spatial distribution of radon decays}
\label{sec:spatialdist}

\begin{figure}[!ht]
 \centering
  \includegraphics[width=\textwidth]{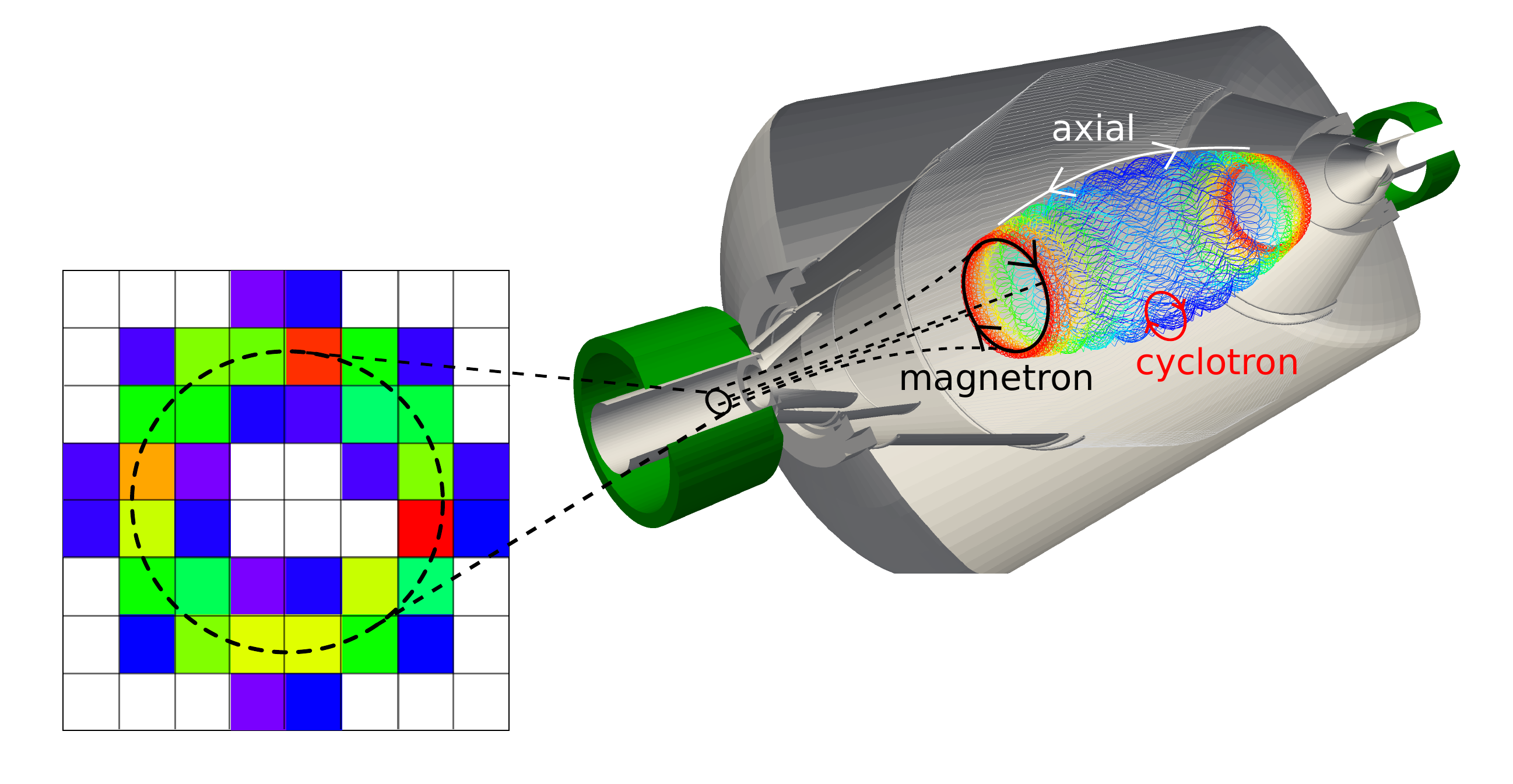}
 \caption{Simulated trajectory of a single trapped electron with start energy $E=1000~\text{eV}$. The electron motion consists of a very fast cyclotron motion around the magnetic field line, a fast axial motion and a slower magnetron motion around the beam axis. Secondary electrons generated by the primary electron along its path are therefore seen as rings on the pixel detector. One can identify the main hit region (green to red colors, corresponding to a large number of hits) and a surrounding fuzzy region (blue, only a few hits) due to the cyclotron motion of the primary electron. The same signature was found within the measurements of~\cite{Fraenkle}, where figure~6 shows some example events. }
 \label{fig:TrappedElectron}
\end{figure}

Apart from generating elevated levels of background over extended periods of time, the event topology of radon-induced background is an important tool to characterize the background-generating mechanism.
The rather complex motion of stored electrons in a magnetic bottle results in a specific topology.
Due to the excellent radial mapping characteristics of a MAC-E filter (see fig.~\ref{fig:TrappedElectron}), radon-induced background at the 8x8 silicon pixel detector will form a generic ring pattern~\cite{PhDFraenkle,PhDMertens}. 
This radon-induced event topology can be understood by first principles of particle motion, as well as by more detailed simulations of electron trajectories in the \prespectrometer{} set-up. 
The electron motion is composed of a fast cyclotron motion around the guiding magnetic field line, an axial motion between the reflection points of the magnetic mirror, and a slow magnetron motion around the beam axis. 
The magnetron motion is caused by the $\vec{E}\times\vec{B}$ and the $\vec{\nabla}|\vec{B}|\times\vec{B}$ drift, which result from the inhomogeneous electric ($\vec{E}$) and magnetic ($\vec{B}$) field configurations inside the spectrometer.
Secondary electrons, originating from ionizing collisions of the stored primary electron with residual gas molecules, thus monitor this motion by following the magnetic field lines when escaping the magnetic mirror trap.
Consequently, they produce a characteristic ring structure at the detector. 
The example in fig.~\ref{fig:TrappedElectron} shows a main hit region (green to red pixels, multiple hits per pixel) which can easily be identified from the surrounding rather fuzzy region (blue pixels, single hits per pixel) which is caused by the cyclotron motion of the primary electron. 
This unique feature of ring-structures allows to make use of a ring-fitting algorithm to unambiguously identify radon-induced background events~\cite{PhDFraenkle}.

% Kenntnis Detektorposition - abhängig?

\begin{figure}[ht!]
 \centering
    \includegraphics[width=0.7\textwidth]{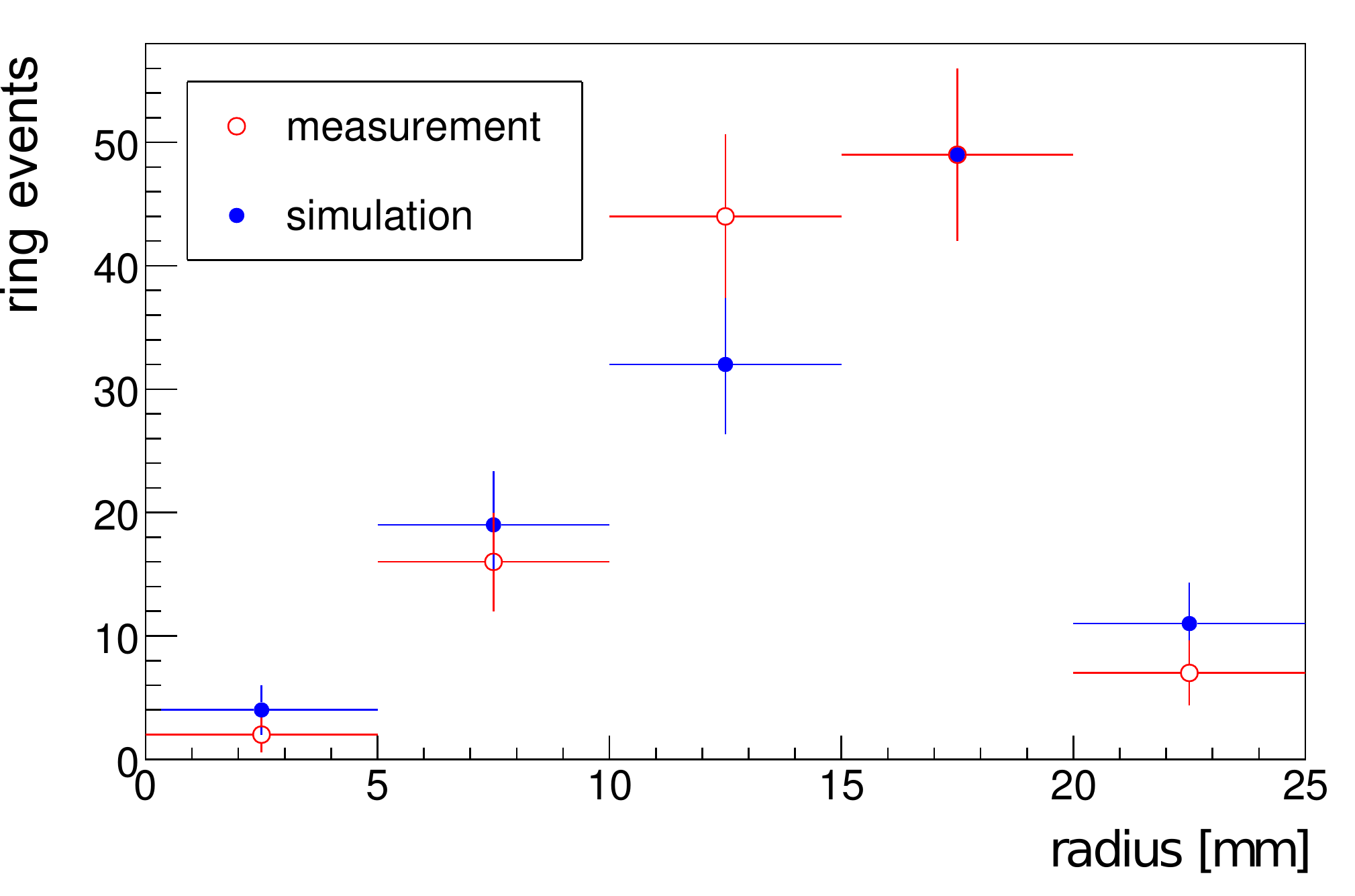}
 \caption{Distribution of fitted ring radii $r_{\text{fit}}$ as determined via measurement and Monte Carlo simulation, and normalized to the total measured event rate. The measured data was adopted from~\cite{Fraenkle}. The good agreement verifies the assumption of a uniform distribution of radon decays inside the spectrometer volume.}
 \label{fig:RingRadius}
\end{figure}

The ring radius fit determines the radial position $r$ of the primary $\alpha$-decay responsible for producing the primary electron relative to the central axis. 
For a homogeneous distribution of $\alpha$-decays inside the spectrometer volume, the number of rings $N(r)$ in a fixed interval $\left[r,r+\text{d}r\right]$ is expected to increase linearly with the radius (see figure~\ref{fig:RingRadius}). 
When comparing measured and simulated spatial ring distributions, the good agreement visible in fig.~\ref{fig:RingRadius} implies that $\alpha$-decays indeed occur with uniform probability over the entire flux tube, as expected for neutral atoms emanating into the UHV region. 
The smaller number of ring structures with radii $r_{\text{fit}}>20$~mm is a result of the limited dimensions of the Si-PIN diode array (length$=40$~mm), which does not cover the entire flux tube (see fig.~\ref{fig:TrappedElectron}).
Nevertheless, events which produce a significant amount of detector hits in the corner pixels with $r_{\text{fit}}>20$~mm can still be identified, albeit with a reduced geometrical efficiency. 

%axial position of the detector -> different magnetic field strength at the detector (e.g. 1cm = xG) and consequently all radii reduced or enhanced by e.g. xG = x mm
% Kenntnis Detektorposition - abhängig?

\subsection{Rate of single events}
\label{sec:ratesingle}

\begin{figure}[ht!]

 \centering
  \subfigure[]{
  \includegraphics[width=0.8\textwidth]{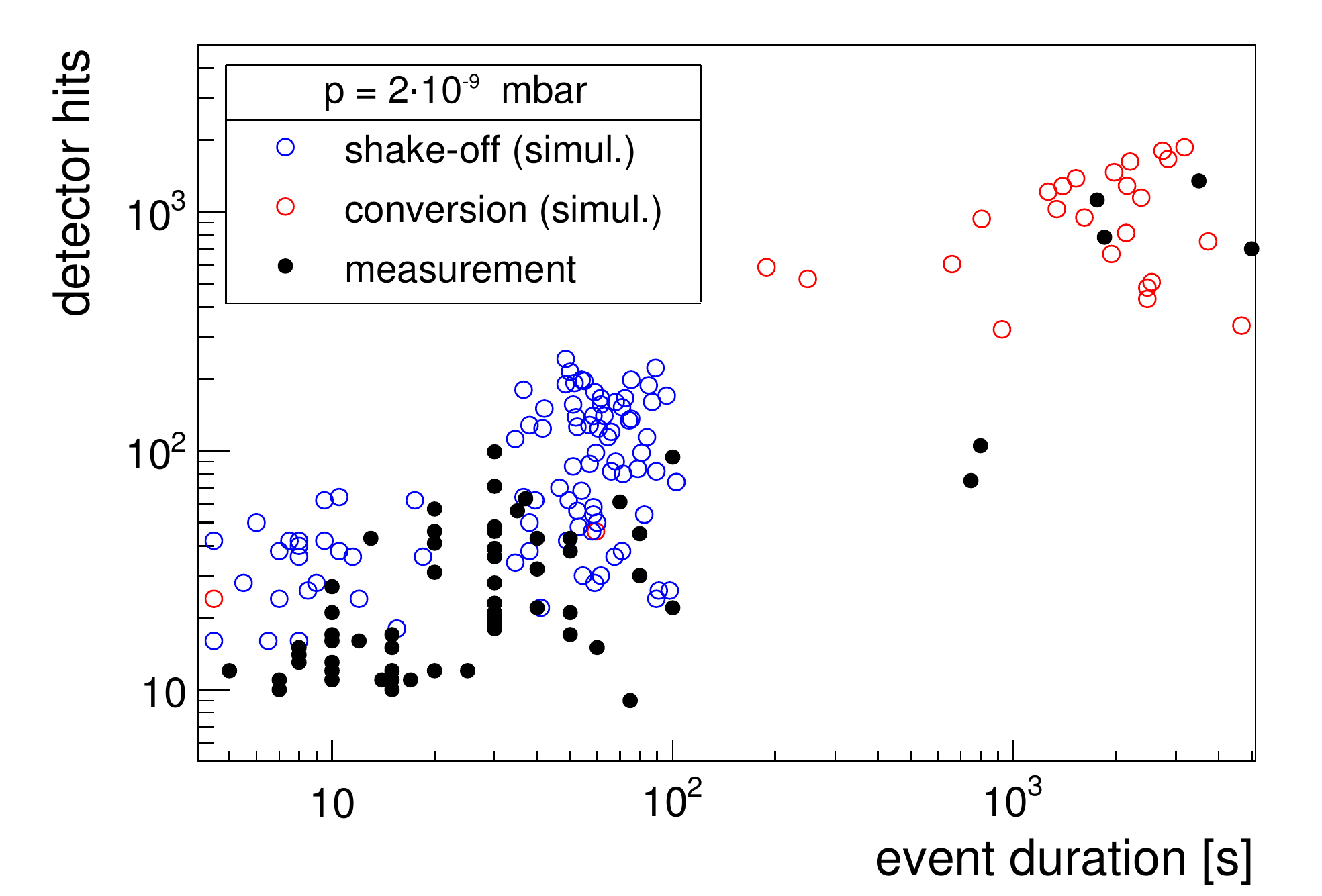}
  }

  \subfigure[]{
  \includegraphics[width=0.8\textwidth]{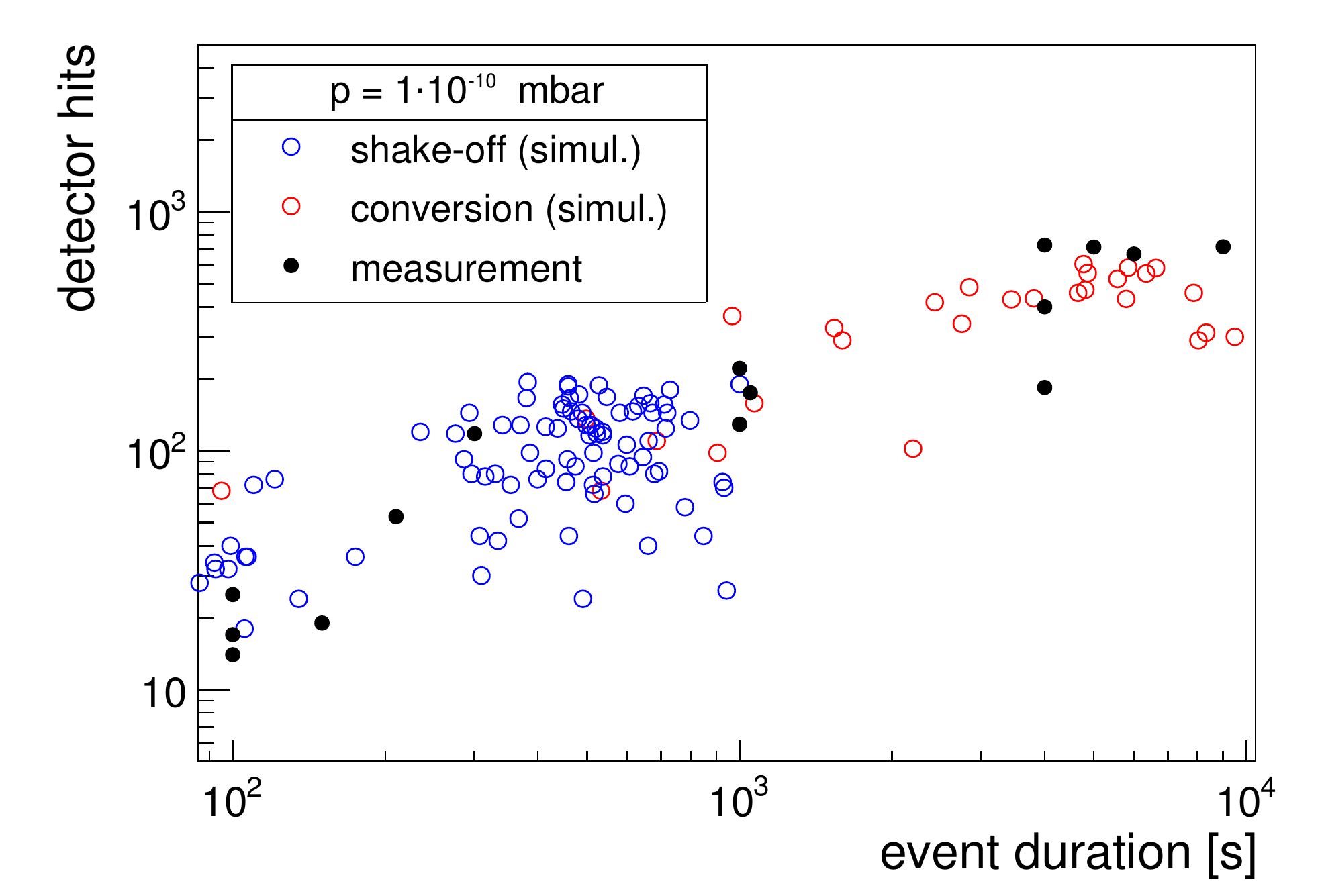}
  }
 \caption{Number of detector hits over event duration for the HPG measurement (a) and the LPG measurement (b). The simulations (open circles) reproduce the features of the measurements (full circles), which is described in more detail in the main text.}
 \label{fig:Rate}

\end{figure}

While the event topology clearly points to a uniform radon decay probability per unit volume over the entire spectrometer volume, we now investigate whether the two other parameters of radon-induced background, namely the number of secondaries and the event duration, also agree with expectations. 
For this investigation we consider two measurements at different pressures (measurements HPG and LPG). 
Figure~\ref{fig:Rate} compares results of measurements and corresponding simulations.
%wieviele Ereignisse in Experiment und Theorie? Wahrscheinlichkeit, dass beide Verteilungen übereinstimmen?

The storage time of a primary electron and the number of secondary electrons it produces strongly depend on the primaries' starting kinetic energy and on the residual gas pressure in the spectrometer volume. 
The number of secondary electrons will increase for higher pressure levels as scattering energy losses increase at the expense of synchrotron energy losses. 
By the same token, the storage time decreases because successive scattering events will happen faster. 
Accordingly, for the HPG measurement, the electron energy losses are dominated by scattering processes (see fig.~\ref{fig:Rate}~(a)). 
Interestingly, both measurement and simulation show two separate, distinct regions with regard to the event duration ($t_{\text{s}}\le10^{2}$~s, $t_{\text{s}}>10^{2}$~s). 
The simulation, which can distinguish between conversion and shake-off events, reveals that this characteristic separation is due to differing primary electron emission processes. 
While the majority of the shake-off electrons has less than 20~keV kinetic energy, conversion electrons typically are found above 100~keV. 
For this pressure regime, the parameter event duration $t_{\text{s}}$ allows to distinguish conversion electrons ($t>10^{2}$~s) from inner shell shake-off processes ($t\le10^{2}$~s). 
On the other hand, at low pressures in the LPG measurement (figure~\ref{fig:Rate}~(b)), synchrotron energy losses tend to smear out this difference. 
While shake-off electrons at low energies are barely affected by synchrotron losses, these losses are dominant in the case of conversion electrons. 
Consequently, the gap between the two emission classes is closed. 
The impact of increased losses due to of synchrotron radiation at excellent UHV conditions is further confirmed by the fact that the overall number of secondary electrons is reduced by a factor of 1.5 for the LPG measurement.

\subsection{Determination of radon activities}
\label{sec:activities}

Following the above considerations, the cooling time of a single electron varies between a few seconds for very low energies at high pressures, and a few hours for the largest energies at low pressures. 
In case of the \prespectrometer{} background measurements the radon activity was low enough so that the time between two events was larger than the event duration. 
Therefore, individual $\alpha$-decays can be clearly discriminated, which allows for their counting.

The excellent agreement between Monte Carlo simulations and experimental data, as well as the different vacuum conditions of the three measurements (LPG, HPG and HP), which influence shake-off and conversion electrons differently, can now be used to determine the $\alpha$-decay activities of the two isotopes (${}^{219}$Rn or ${}^{220}$Rn), as well as their origin (getter, bulk) yielding the four observables ${}^{219}\text{Rn}_{\text{B}},\,{}^{219}\text{Rn}_{\text{G}},\,{}^{220}\text{Rn}_{\text{B}}$ and ${}^{220}\text{Rn}_{\text{G}}$. 
To discriminate between ${}^{219}$Rn and ${}^{220}$Rn induced events, we use the simulated decay probabilities for CI-III events to fit the experimental data of table~\ref{tab:measurement}. 
We finally compare the more detailed results of this work to our earlier results in~\cite{Fraenkle}, which were based on measurements only.

Table~\ref{tab:MCResults} summarizes the simulated probabilities of CI-CIII events following ${}^{219}$Rn and ${}^{220}$Rn decays for the three experimental configurations. 
The key experimental parameters pressure and gas composition are identical to table~\ref{tab:measurement}. 
Furthermore, the average number of detector hits per event $\left<N_{\text{det}}\right>$ is shown, which is required to determine the actual background contribution.

\begin{table}[!ht]
\caption{Overview of simulation results, comprising 10000 electrons for each configuration and radon isotope. 
 The probability $P$ for the occurrence of CI-III events per decay and the average number of detector hits $\left<N_{det}\right>$ per event are shown.}
\begin{center}
  \begin{tabular*}{0.92\textwidth}{p{4.2cm} p{2.6cm} p{2.6cm} p{2.7cm}}
  \hline
  \hline
  measurement & LPG & HPG & HP\\
  \hline
  \end{tabular*}
  \begin{tabular}{l|cc|cc|cc}
  radon type & $^{219}$Rn & $^{220}$Rn & $^{219}$Rn & $^{220}$Rn & $^{219}$Rn & $^{220}$Rn\\
  \hline
  $P$ [$10^{-3}$] (CI) &  8  & 5.8  &  9.2  & 5  &  6.8 & 5.7\\
  $\left<N_{\text{det}}\right>$/event (CI) & 22.8 & 19  &   24.3 & 21.9  &  25.3 & 21.4 \\
  \hline
  $P$ [$10^{-3}$] (CII)& 3.9 & 2.1 &  3.3 & 0.8 & 3.3 & 0.8\\
  $\left<N_{\text{det}}\right>$/event (CII) & 130.3 & 71.6 & 123.2 & 58.6 & 134.4 & 51.3 \\
  \hline
  $P$ [$10^{-3}$] (CIII) & 4.2 & 0 & 2.7 & 0 & 5.3 & 0\\
  $\left<N_{\text{det}}\right>$/event (CIII) & 677.9 & 0 & 932.7 & 0 & 1033.8 & 0 \\
  \hline
  \hline
  \end{tabular} 
\end{center}
\label{tab:MCResults}
\end{table}

\begin{figure}[ht!]
 \centering
  \includegraphics[width=0.7\textwidth]{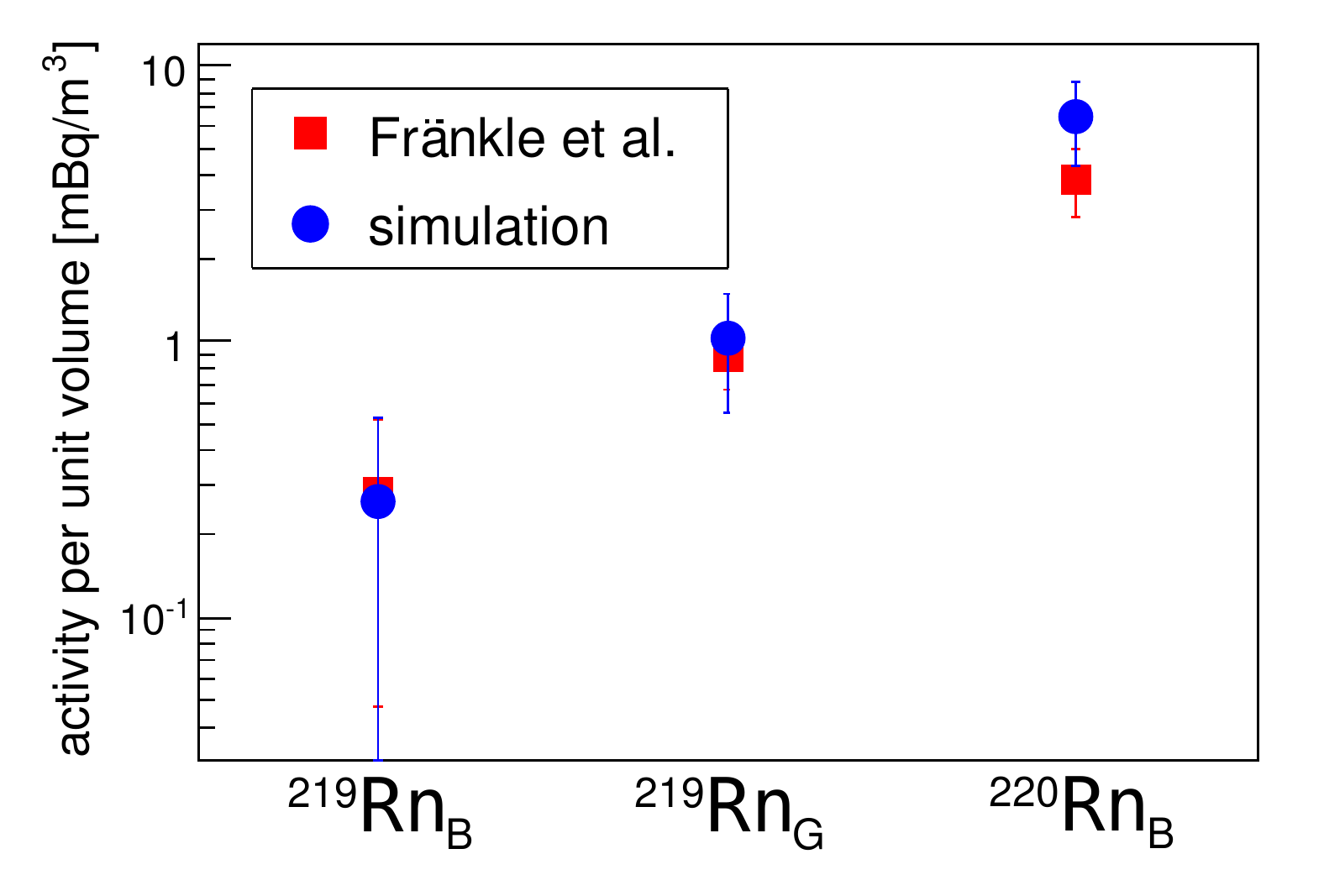}
 \caption{Total activity of ${}^{219}\mathrm{Rn}_{\text{B}}$ (bulk material of the spectrometer vessel), ${}^{219}\mathrm{Rn}_{\text{G}}$ (getter material) and ${}^{220}\mathrm{Rn}_{\text{B}}$ inside the \prespectrometer{}. The values have been determined by a three-parameter fit of simulated to measured event rates. The simulation results of this work (circles) are compared to values derived from measurements of Fr{\"a}nkle~et~al.~\cite{Fraenkle} (squares).}
 \label{fig:activity}
\end{figure}

The event rates $r_{i}$ for the individual classes $C_{i}$, with $i=I,II,III$, are determined from the activities of the three different radon sources $A({}^{219}\text{Rn}_{\text{B}})$, $A({}^{219}\text{Rn}_{\text{G}})$ and $A({}^{220}\text{Rn}_{\text{B}})$, the corresponding probabilities $P_{i}$ for the occurrence of an event of class $C_{i}$ and the decay probability $\epsilon$:
\begin{equation}
 r_{i}=\sum_{k={}^{219}\text{Rn}_{\text{B}},{}^{219}\text{Rn}_{\text{G}},{}^{220}\text{Rn}_{\text{B}}}\epsilon(k)\cdot A(k)\cdot P_{i}(k).
\label{equ:eventRate}
\end{equation}
The probabilities $P_{i}(k)$ are taken from table~\ref{tab:MCResults} and the decay probabilities $\epsilon(k)$ from table~\ref{tab:decayProb}. 
The only free parameters remaining are thus the radon activities $A(k)$, which can be determined by a three-parameter $\chi^2$-fit of the simulated event rates $r_{i}$ to the measured rates of table~\ref{tab:measurement}. 
In fig.~\ref{fig:activity} we show the fit results for the radon activities per unit volume in the \prespectrometer{} (total volume: 8.5~$\text{m}^{3}$) and compare them to the activities which were observed in the measurements of Fr{\"a}nkle~et~al.~\cite{Fraenkle}. 
The simulated activities in general are somewhat larger than the measured ones, which can be explained by two facts. 
First, the effects of non-adiabaticity were not considered in~\cite{Fraenkle}.
Furthermore, our extensive simulations have revealed that some CI events do not appear as rings on the detector, and, consequently, could not be attributed to radon-induced background within the analysis of~\cite{Fraenkle}.

Table~\ref{tab:GerdaComparison} gives the emanation rates of ${}^{219,220}\mathrm{Rn}_{\text{B}}$ into the KATRIN \prespectrometer{} stainless steel vessel per unit volume.
The values are compared to the independent measurement of the ${}^{222}\mathrm{Rn}$ emanation for the empty and fully equipped GERDA cryostat, as reported in~\cite{Simgen,SimgenCollab}.
Unfortunately, the measurement technique applied in~\cite{Simgen} does not allow detection of the short-lived ${}^{219,220}\mathrm{Rn}$ isotopes. 
As the authors of~\cite{Simgen} point out, stainless steel vessels show much larger radon emanation rates than pure stainless steel samples, which can be caused by surface impurities, in particular due to welding procedures. 
Furthermore, any auxiliary equipment attached to the vessel will significantly increase the radon emanation rate, an effect both observed in KATRIN and GERDA for different isotopes. 

\begin{table}[!ht]
\caption{Comparison of radon emanation rates per unit volume in stainless steel vessels. The simulated ${}^{219,220}\mathrm{Rn}_{\text{B}}$ concentrations for the fully equipped KATRIN \prespectrometer{} stainless steel vessel ($V=8.5~\text{m}^{3}$, $A=25~\text{m}^{2}$) are compared to the ${}^{222}\mathrm{Rn}$ concentrations in case of an empty and a fully equipped GERDA cryostat~\cite{Simgen,SimgenCollab} ($V=65~\text{m}^{3}$, $A=70~\text{m}^{2}$).}
\begin{center}
  \begin{tabular}{lc}
  \hline
  \hline
   & concentration [mBq/$\text{m}^{3}$] \\
  \hline
   ${}^{219}\mathrm{Rn}$ [this work] & $0.26\pm0.26$\\
   ${}^{220}\mathrm{Rn}$ [this work] & $6.53\pm2.18$\\
   ${}^{222}\mathrm{Rn}$~\cite{Simgen} (empty) & $0.22\pm0.03$ \\
   ${}^{222}\mathrm{Rn}$~\cite{SimgenCollab} (fully equipped) & $0.85\pm0.06$ \\
  \hline
  \hline
  \end{tabular} 
\end{center}
\label{tab:GerdaComparison}
\end{table}

\begin{figure}[ht!]
 \centering
  \includegraphics[width=0.75\textwidth]{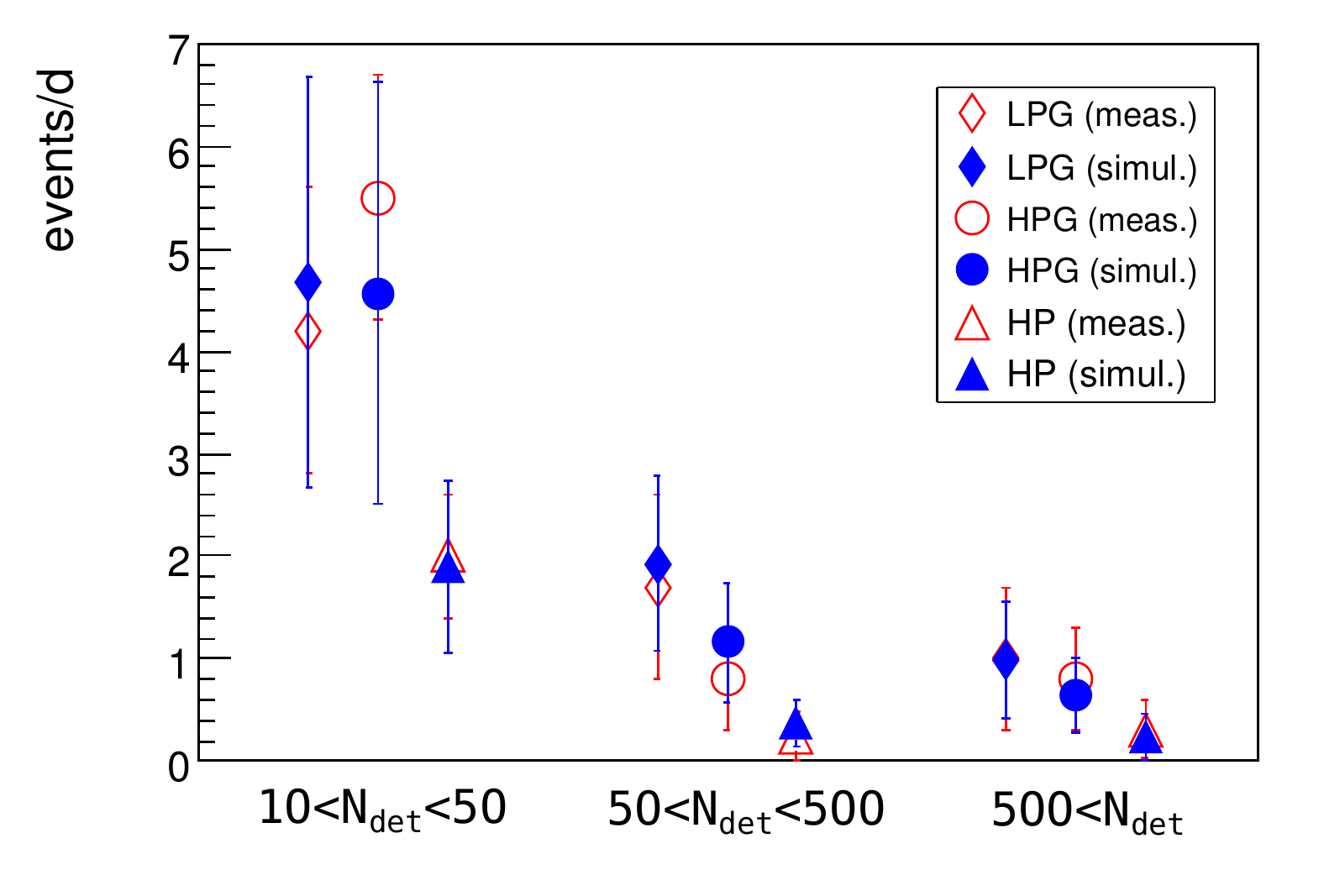}
  \caption{Event rates for the individual classes and measurements, determined according to eq.(\ref{equ:eventRate}). The simulations (full symbols, blue) are in good agreement with the measurement results (open symbols, red).}
 \label{fig:FitResult}
\end{figure}

Figure~\ref{fig:FitResult} compares the event rates determined according to eq.(\ref{equ:eventRate}) to those derived within the measurements of this work. 
The values are in good agreement within their errors (propagated from the errors on the activities).

%Andere Auftragung (x-Achse und Legende tauschen)

\begin{figure}[ht!]
 \centering
  \includegraphics[width=0.75\textwidth]{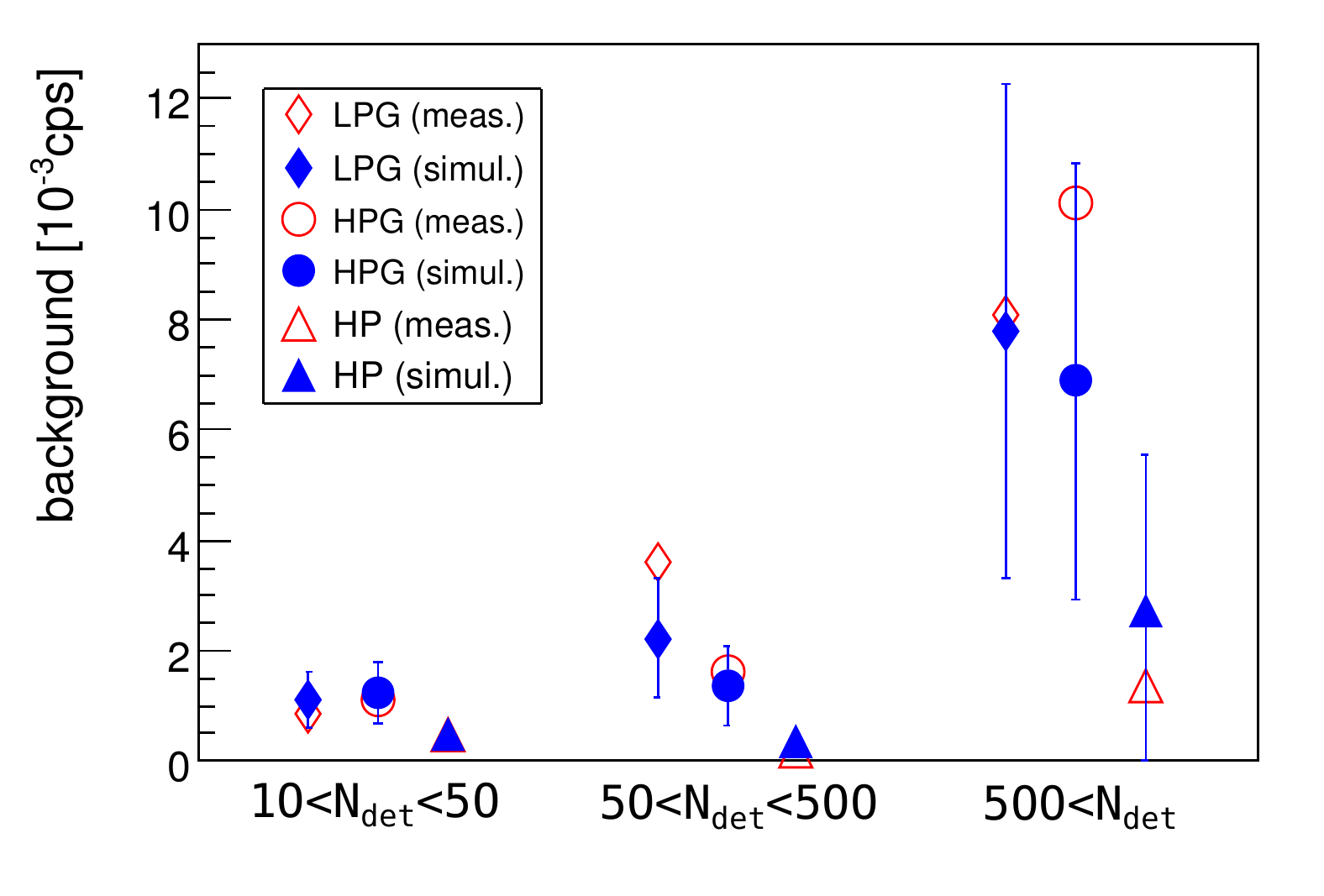}
 \caption{Background contribution from the individual classes. The simulations (full symbols, blue) are in agreement with the measurement results (open symbols, red).}
 \label{fig:BackgroundContribution}
\end{figure}

In figure~\ref{fig:BackgroundContribution}, the background contributions of the individual classes are shown. 
These values are determined by multiplying the calculated event rates with the simulated average number of detector hits per event $\left<N_{\text{hit,MC}}\right>$ of each class for the different radon isotopes.

After subtracting the contributions of CI-III events from the total measured background rate, a background component of about $3\cdot10^{-3}$~cps remains. 
A fraction of this background results from radon decays which produce less than 10 detector hits, or are not detected as ring events by the analysis software (in the following labeled C0 events). 
These are mainly $\alpha$-decay events where only two low-energy shell reorganization electrons are emitted~\cite{RadonModel}, which have a low probability of being magnetically stored in the \prespectrometer{}. 
The simulations reveal that these radon decays produce on average 0.2 detector hits. 
Table~\ref{tab:C0Events} summarizes the C0 contributions to the background rate for the three measurements considered. 
The fact that all simulated rates contribute significantly, but do not exceed the remaining measured single hit background rate is another important validation of our background model, showing that radon-induced processes also contribute to the measured C0 class (possibly saturating it).

\begin{table}[!ht]
\caption{Simulated and measured background rate ($r_{\mathrm{C0,simu}},\;r_{\mathrm{C0,meas}}$) due to C0 single hit events.}
\begin{center}
  \begin{tabular}{lcc}
  \hline
  \hline
  measurement & $r_{\mathrm{C0,simu}}$ [$10^{-3}$~cps]  &$r_{\mathrm{C0,meas}}$ [$10^{-3}$~cps] \\
  \hline
  LPG  & $1.6\pm0.03$ & $3.2\pm0.3$\\
  HPG  & $1.4\pm0.03$ & $3.4\pm0.3$\\
  HP   & $0.8\pm0.02$ & $2.0\pm0.3$\\
  \hline
  \hline
  \end{tabular} 
\end{center}
\label{tab:C0Events}
\end{table}

%==============================================================================
%==============================================================================

%Conclusion
\section{Conclusions}

In the course of this work we have developed a detailed model of electron emission processes following the $\alpha$-decays of the two radon isotopes ${}^{219}$Rn and ${}^{220}$Rn. 
These investigations were motivated by our earlier observations, reported in~\cite{Fraenkle}, of periods with significantly enhanced background rates at the KATRIN \prespectrometer{} measurements.

The background model incorporates various processes which lead to the emission of electrons such as internal conversion, inner shell shake-off and relaxation of the atomic shells during or after the $\alpha$-emission. 
The radon event generator described in~\cite{RadonModel} has been used as input for extensive Monte Carlo simulations with the \textsc{Kassiopeia} simulation package. 
The validity of our background model and corresponding Monte Carlo simulations has been confirmed by a comparison with key experimental observables such as event duration and number of detector hits. 
The relative contribution of the two isotopes ${}^{219}$Rn and ${}^{220}$Rn has been determined by varying the pressure in the UHV recipient. 
In addition, by removing the NEG strips from the \prespectrometer{} pump port, we were able to assess the contributions from the spectrometer surface. 
As a result, the radon-induced background has been fully characterized. 
An important outcome of these investigations is the realization that low-energy shell reorganization electrons comprise a significant fraction of the single hit background rate.

These findings are of major importance for the upcoming KATRIN measurements with the main spectrometer. 
In~\cite{NuclearDecay} we extrapolated the background model of this work to the different electromagnetic layout (minimum magnetic field 3~mT instead of 156~mT here) at the large main spectrometer, taking into account also the much larger NEG pump in operation there. 
The work presented in this publication has also been instrumental in developing active~\cite{ECR} as well as passive~\cite{Baffle} countermeasures against trapped electrons following radon $\alpha$-decays.

It is only by developing and by validating detailed models of background processes that the KATRIN experiment can realize its full physics potential in measuring the absolute mass scale of neutrinos.

%Acknowledgements
\section*{Acknowledgement}
 
This work has been supported by the Bundesministerium f{\"u}r Bildung und Forschung (BMBF) with project number 05A08VK2 and the Deutsche Forschungsgemeinschaft (DFG) via Transregio 27 ``Neutrinos and beyond''. 
We also would like to thank the Karlsruhe House of Young Scientists (KHYS) of KIT for their support (S.G., S.M., N.W.).

%==============================================================================
%==============================================================================

%References

\bibliographystyle{unsrt}

\section*{References}

\begin{thebibliography}{10}
\expandafter\ifx\csname url\endcsname\relax
  \def\url#1{\texttt{#1}}\fi
\expandafter\ifx\csname urlprefix\endcsname\relax\def\urlprefix{URL }\fi
\expandafter\ifx\csname href\endcsname\relax
  \def\href#1#2{#2} \def\path#1{#1}\fi

\bibitem{NeutrinoOscillation}
Y.~Fukuda {\em et~al.}, ``Evidence for oscillation of atmospheric neutrinos,''
  {\em Phys. Rev. Lett.}, vol.~81, pp.~1562--1567, 1998.

\bibitem{King}
S.F.~King, ``Neutrino Mass,''
  {\em 	arXiv:0712.1750v1}, 2007.

\bibitem{m_nu_cosmo}
J.~Lesgourguesa and S.~Pastor, ``Massive neutrinos and cosmology,''
  {\em Phys. Rep.}, vol.~429, nr.~6, pp.~307--379, 2006.

\bibitem{MassHierarchy}
N.~K.~Francis and N.~N.~Singh, ``{Quasi-Degenerate Neutrino Masses with Normal and Inverted Hierarchy},'' 
  {\em J. Mod. Phys.}, vol.~2 pp.~1280--1284, 2011.

\bibitem{Cosmo}
S.~Hannestad, ``{Neutrino physics from precision cosmology},'' 
  {\em Prog. Part. Nucl. Phys.}, vol.~65 pp.~185--208, 2010.

\bibitem{EXO}
J.J.~Gomez-Cadenas {\em et~al.}, ``{The search for neutrinoless double beta decay},'' 
  {\em Riv. Nuovo Cim.}, vol.~35, pp.~29--98, 2012.

\bibitem{0NBB}
A.S.~Barabash, ``{Double Beta Decay Experiments},'' 
  {\em Phys. Part. Nucl.}, vol.~42, no.~4, pp.~613--627, 2011.

\bibitem{MAHO}
M.~Galeazzi {\em et~al.}, ``{The Electron Capture Decay of 163-Ho to Measure the Electron Neutrino Mass with sub-eV Accuracy (and Beyond)},'' 
  {\em 	arXiv:1202.4763v2}, 2012.

\bibitem{Review}
G.~Drexlin {\em et~al.}, ``{Current Direct Neutrino Mass Experiments},'' 
  {\em Adv. High  }, 2013.

\bibitem{Mainz}
C.~Kraus {\em et~al.}, ``{Final Results from phase II of the Mainz Neutrino Mass Search in Tritium $\beta$ Decay},'' 
  {\em Eur. Phys. J. C}, vol.~40, pp.~447--468, April 2005.

\bibitem{DesignReport}
``{KATRIN Design Report (FZKA Report 7090)},'' tech. rep.,\url{http://www.katrin.kit.edu/}, KIT, 2004.

\bibitem{WGTS}
W.~K{\"a}fer, ``{The Windowless Gaseous Tritium Source of KATRIN},'' 
  {\em Prog. Part. Nucl. Phys.}, vol.~64, pp.~297--299, April 2010.

\bibitem{MACE}
G.~Beamson, H.~Q. Porter, and D.~W. Turner, ``{The collimating and magnifying properties of a superconductiong field photoelectron spectrometer},'' 
  {\em J.  Phys. E}, vol.~13, no.~64, 1980.

\bibitem{MACE1}
V.~M. Lobashev and P.~E. Spivak, ``{A method for measuring the electron antineutrino rest mass},''
  {\em Nucl. Instrum. Meth. A}, vol.~240, no.~2, pp.~305--310, 1985.

\bibitem{MACE2}
A.~Picard {\em et~al.}, ``{A solenoid retarding spectrometer with high resolution and transmission for keV electrons},'' 
  {\em Nucl. Instrum. Meth. B}, vol.~63, no.~3, pp.~345--358, 1992.

\bibitem{Fraenkle}
F.~Fr{\"a}nkle {\em et~al.}, ``{Radon induced background processes in the KATRIN pre-spectrometer},'' 
  {\em Astropart. Phys.}, vol.~35, no.~3, pp.~128--134, 2011.

\bibitem{RadonModel}
N.~Wandkowsky {\em et~al.}, ``Modeling of electron emission processes accompanying Radon-$\alpha$-decays within electrostatic spectrometers,''
  {\em submitted to J. Phys. G}, 2013.

\bibitem{MagneticMirror}
H.~Higaki, K.~Ito, K.~Kira, and H.~Okamoto, ``{Electrons Confined with an Axially Symmetric Magnetic Mirror Field},'' 
  {\em AIP Conference Proceedings}, vol.~1037, no.~1, pp.~106--114, 2008.

\bibitem{NuclearDecay}
S.~Mertens {\em et~al.}, ``{Background due to stored electrons following nuclear decays at the KATRIN experiment},'' 
  {\em Astropart. Phys.} vol.~41, no.~52, 2013.

\bibitem{Vacuum}
J.~Wolf, ``Size matters: The vacuum system of the Katrin neutrino experiment,''
  {\em Journal of the Vacuum Society of Japan}, vol.~52, pp.~278--284, 2009.

\bibitem{PhDFraenkle}
{F.~Fr{\"a}nkle}, {\em {Background Investigations of the KATRIN Pre-Spectrometer}}.
\newblock PhD thesis, Karlsruhe Institute of Technologie (KIT), 2010.

\bibitem{PhDMertens}
{S.~Mertens}, {\em {Study of background processes in the electrostatic spectrometers of the KATRIN experiment}}.
\newblock PhD thesis, Karlsruhe Institute of Technologie (KIT), 2012.

\bibitem{PhDNancy}
{N.~Wandkowsky}, {\em {PhD thesis in preparation}}.%{\em {Study of background processes in the electrostatic spectrometers of the KATRIN experiment}}.
\newblock Karlsruhe Institute of Technologie (KIT), 2013.

\bibitem{ECR}
S.~Mertens {\em et~al.}, ``{Stochastic Heating by ECR as a Novel Means of Background Reduction in the KATRIN spectrometers},'' 
  {\em JINST} 7 P08025, 2012.

\bibitem{Kassiopeia}
D.~Furse {\em et~al.}, ``{KASSIOPEIA - the simulation package for the KATRIN experiment},''
\newblock to be published.

\bibitem{ShakeOff}
M.~S. Freedman, ``{Ionization by Nuclear Transitions},''
 {\em Summer course in atomic physics, Carry-le-Rouet, France}, Aug 1975.

\bibitem{KShakeOffRadon}
M.~S. Rapaport, F.~Asaro, and I.~Perlman, ``{$K$-shell electron shake-off accompanying alpha decay},'' 
  {\em Phys. Rev. C}, vol.~11, pp.~1740--1745, May 1975.

\bibitem{LMShakeOffRadon}
M.~S. Rapaport, F.~Asaro, and I.~Perlman, ``{$M$- and $L$-shell electron shake-off accompanying alpha decay},'' 
  {\em Phys. Rev. C}, vol.~11, pp.~1746--1754, May 1975.

\bibitem{Bang}
J.~Bang and J.~M. Hansteen, ``{Coulomb deflection effects on ionization and pair-production phenomena},'' 
  {\em K. Dan. Vidensk. Selsk. Mat. - Fys. Medd.}, vol.~31, no.~13, pp.~1--43, 1959.

\bibitem{ConversionDataRn219}
E.~Browne, ``{Nuclear Data Sheets for $A = 215, 219, 223, 227, 231$},'' 
  {\em Nuclear Data Sheets}, vol.~93, no.~4, pp.~763--1061, 2001.

\bibitem{ConversionDataRn220}
S.-C. Wu, ``{Nuclear Data Sheets for $A = 216$},'' 
  {\em Nuclear Data Sheets}, vol.~108, no.~5, pp.~1057--1092, 2007.

\bibitem{Larkins}
F.~P.~Larkins, ``{Semiempricial Auger-electron energies for elements $10\leq Z\leq 100$},'' 
  {\em At. Data Nucl. Data Tables}, vol.~20, no.~4, pp.~311--387, 1977.

\bibitem{Pomplun}
E.~Pomplun, ``{Auger Electron Spectra - The Basic Data for Understanding the Auger Effect},'' 
  {\em Acta Oncologica}, vol.~39, no.~6, pp.~673--679, 2000.

\bibitem{ShellReorganization}
J.~S. Hansen, ``{Internal ionization during alpha decay: A new theoretical approach},'' 
  {\em Phys. Rev. A}, vol.~9, pp.~40--43, Jan 1974.

\bibitem{ShellReorg}
M.~S. Freedman, ``{Atomic structure effects in nuclear events},'' 
  {\em Annu. Rev. Nucl. Sci.}, vol.~24.

\bibitem{RungeKutta1}
J.~H. Verner, ``{Explicit Runge-Kutta methods with estimates of the local truncation error},'' 
  {\em SIAM J. Numer. Anal.}, vol.~15, pp.~772--290, 1978.

\bibitem{RungeKutta2}
P.~Prince and J.~Dormand, ``{High order embedded Runge-Kutta formulae},'' 
  {\em J. Comput. Appl. Math.}, vol.~7, no.~1, pp.~67--75, 1981.

\bibitem{RungeKutta3}
C.~Tsitouras and S.~N. Papakostas, ``{Cheap error estimation for runge--kutta methods},'' 
  {\em SIAM J. Sci. Comp.}, vol.~20, no.~6, pp.~2067--2088, 1999.

\bibitem{Electrode}
K.~Valerius, ``{The wire electrode system for the KATRIN main spectrometer},'' 
  {\em Prog. Part. Nucl. Phys.}, vol.~64, no.~2, pp.~291--293, 2010.

\bibitem{Aircoil}
F.~Gl{\"u}ck {\em et~al.}, ``{Electromagnetic design of the KATRIN large volume air coil system},'' 
  {}, to be published, 2013.

\bibitem{FerencEl}
F.~Gl{\"u}ck, ``{Axisymmetric electric field calculation with zonal harmonic expansion},'' 
  {\em Progress In Electromagnetics Research B}, vol.~32, pp.~319--350, 2011.

\bibitem{FerencMag}
F.~Gl{\"u}ck, ``{Axisymmetric magnetic field calculation with zonal harmonic expansion},'' 
  {\em Progress In Electromagnetics Research B}, vol.~32, pp.~351--388, 2011.

\bibitem{BEM}
P.~W. Hawkes and E.~Kasper, {\em {Principles of Electron Optics}}, vol.~1.
\newblock Academic Press, 1989.

\bibitem{ElasticHydrogen1}
J.~Liu and S.~Hagstrom, ``{Dissociative cross section of $H_{2}$ by electron impact},'' 
{\em {Phys. Rev. A}}, vol.~50, no.~4, 1994.

\bibitem{ElasticHydrogen2}
S.~Trajmar and D.~F.~Register and A.~Chutjian, ``{Electron scattering by molecules II. Experimental methods and data},'' 
{\em {Phys. Rep.}}, vol.~97, pp.~219--356, 1983.

\bibitem{BEB}
Y.-K.~Kim and M.~E.~Rudd, ``{Binary-encounter-dipole model for electron impact ionization},'' 
{\em {Phys. Rev. A}}, vol.~50, no.~5, 1994.

\bibitem{BEB2}
W.~Hwang and Y.-K.~Kim and M.~E.~Rudd, ``{New model for electron-impact ionization cross section of molecules},'' 
{\em {J. Chem. Phys.}}, vol.~104, no.~8, 1996.

\bibitem{ExcitationHydrogen}
G.~Arrighini, F.~Biondi, and C.~Guidotti, ``{A study of the inelastic scattering of fast electrons from molecular hydrogen},'' 
  {\em Mol. Phys.}, vol.~41, no.~6, pp.~1501--1514, 1994.

\bibitem{ExcitationHydrogen2}
Z.~Chen and A.~Z.~Msezane, ``{Calculation of the excitation cross sections for the ${}^{1}\Sigma_{u}^{+}$ and $C{}^{1}\Pi_{u}^{+}$ states in $e-H_{2}$ scattering at 60~eV},'' 
  {\em Phys. Rev. A}, vol.~51, no.~5, pp.~3745--3750, 1995.

\bibitem{Molecules}
E.~Gargioni and B.~Grosswendt, ``{Electron-Impact Cross Sections for Ionization and Excitation},'' 
  {\url{http://physics.nist.gov/PhysRefData/Ionization/molTable.html}}.


\bibitem{Argon}
E.~Gargioni and B.~Grosswendt, ``{Electron scattering from argon: Data evaluation and consistency},'' 
  {\em Rev. Mod. Phys.}, vol.~80, pp.~451--480, 2008.

\bibitem{Gerda}
W.~Maneschg {\em et~al.}, ``{Measurement of extremely low radioactivity levels in stainless steel for GERDA},'' 
  {\em Nucl. Inst. Meth. A}, vol.~593, pp.~448--453, 2008.

\bibitem{NEG}
X.~Luo, L.~Bornschein, C.~Day, and J.~Wolf, ``{KATRIN NEG pumping concept investigation},'' 
  {\em Proceedings of the European Vacuum Conference (EVC-9)}, vol.~81, no.~6, pp.~777--781, 2007.

\bibitem{Eloss}
S.~P. Khare, ``{Ionizing Collisions of Electrons with Atoms and Molecules},''
  {\em Radiation Research}, vol.~64, no.~1, pp.~106--118, 1975.

\bibitem{Simgen}
G.~Zuzel and H.~Simgen, ``{High sensitivity radon emanation measurements},'' 
  {\em Appl. Rad. Isot.}, vol.~67, no.~5, pp.~889-893, 2009.

\bibitem{SimgenCollab}
H.~Simgen, ``{Radon Background in Low Level Experiments},'' 
  \newblock talk given at the {\em XIX. KATRIN Collaboration meeting}, 2010.

\bibitem{Baffle}
S.~G{\"o}rhardt {\em et~al.}, ``{Combination of baffle and cold trap for radon trapping in the KATRIN spectrometers},'' 
  to be published, 2013.


\end{thebibliography}

\end{document}